# Measurements of 10 Scarcely Observed Pairs

Matthew B. James[1], Graeme L. White[2], Roderick R. Letchford[2], Stephen G. Bosi[1].

1) University of New England, NSW, Australia.
   m.b.james27@gmail.com; sbosi@une.edu.au
2) University of Southern Queensland, QLD, Australia.
   graemewhiteau@gmail.com; Rod.Letchford@usq.edu.au

**Abstract.**

Separation ($\rho$) and Position Angle (*PA*) measurements are reported of 10 pairs which measures where last reported in the WDS +20 years from epoch of observation 2021.066. Measurements were obtained by direct imaging and are presented with associated measurement uncertainties, as well as, comparisons to measurements determined from Gaia DR2 & EDR3 and historic data extrapolation at epoch of J2000.0.

## 1.0 Observations.

*Instrumentation*

The 14-inch Schmidt-Cassegrain located at the Kirby observatory owned by the University of New England was used to make the observations in this work. The camera used for the observation was a ZWO ASI-120MM-S monochrome CMOS camera fitted with a Wratten #25 red filter.

*Software & Data Acquisition*

The capture software Sharpcaps (Glover) was used for the observations. Observational data was processed in the *Reduc* software. The pixel scale was determined by reference pair calibration using direct imaging of αCen AB in *Reduc* which produced a 0.1766 as/px scale; the same calibration used previous works (James, 2019 & 2020a). The camera rotation was computed using the *Synthetic Drift* function. The *AutoReduc* function was then used and all images greater than 2 standard deviations from the mean of *PA* and $\rho$ were rejected from this work. The *Aladin Sky Atlas* and the *SIMBAD* astronomical database was used to source the Gaia data.

*Selection of Pairs*

The 10 pairs in this work were sourced from the Washington Double Star (WDS) Catalog with the requirements that the pairs had a last recorded separation greater than 4 arcseconds, a last recorded observation greater than 20 years, that the secondary was brighter than magnitude 10 and that the delta magnitude was less than 2.5. The exceptions were the close pair 06387-4504 HJ 3882AB which was detectable observing 06387-4504 HJ 3882AC, and 08057-3334 HJ 4046AC ($\Delta$mag ≈ (4.39,4.74); Gaia Collaboration, *et al.*) as it was presented in a private communication (Damm, F., 2018).

## 2.0 Astrometric Measures

This work was conducted on epoch of observation 2021.066. All pairs, excluding 06387-4504 HJ 3882AC (1991) and 08057-3334 HJ 4046AC (2015), were last updated in the WDS in

1999. Table 1 presents the measurements of *PA* and *ρ* with their associated standard error in the mean (SEM).

Table 1: *PA* and *ρ* determined from observations at epoch 2021.066.

| # | WDS | DISC | PA (deg) | PA SEM (deg) | ρ (arcsec) | ρ SEM (arcsec) |
|---|---|---|---|---|---|---|
| 1 | 06018-4110 | HJ 3827AB,C | 239.86 | 0.007 | 23.844 | 0.003 |
| 2 | 06387-4504 | HJ 3882AB | 28.23 | 0.252 | 1.942 | 0.011 |
| 3 | 06387-4504 | HJ 3882AC | 331.53 | 0.076 | 17.992 | 0.020 |
| 4 | 06393-3150 | VOU 21A,BC | 27.81 | 0.018 | 23.752 | 0.008 |
| 5 | 06443-3219 | COO 43 | 264.52 | 0.071 | 4.719 | 0.007 |
| 6 | 07289-3151 | DUN 49 | 53.89 | 0.029 | 8.932 | 0.007 |
| 7 | 07341-5050 | HJ 3986 | 220.08 | 0.013 | 49.517 | 0.013 |
| 8 | 08057-3334 | HJ 4046AB | 88.43 | 0.017 | 22.061 | 0.007 |
| 9 | 08057-3334 | HJ 4046AC | 60.42 | 0.062 | 27.911 | 0.028 |
| 10 | 08057-3334 | I 189BC | 10.80 | 0.371 | 13.285 | 0.078 |

## 3.0 Extrapolated Historical Measurements.

*PA* and *ρ* measures were determined by linear extrapolation from historic observations (sourced from Brian Mason of the UNSO) which also included *PA* and *ρ* measurements from this work and Gaia at epoch J2000.0. Precession based on the proper motion of the historical observation to J2000.0 was performed.

Based on the size of separation, relatively small proper motion, and few reported observations, a linear extrapolation was justified. As an aside, extrapolation of pairs from non-linear curve fitting with smaller separations and larger proper motions can be more accurate, especially over centuries of measurements.

The first listed historic observations of the pairs 06018-4110 HJ 3827AB,C, 06387-4504 HJ 3882AC, 06443-3219 COO 43, 08057-3334 I 189BC were obvious outliers and not used to compute the measures or uncertainties in this work.

Historical uncertainties were estimated from (White, Letchford, & Ernest, 2018), which estimates the accuracy of *PA* and *ρ* measurements of Alpha Centarui AB (αCen AB) over the past ~270 years. αCen AB is a well observed and modelled pair, and with this knowledge, the historical uncertainties presented here are likely smaller than the actual uncertainties of each historical *PA* and *ρ* measurement.

Table 2: *PA* and *ρ* linearly extrapolated to epoch of J2000.0. Extrapolation includes historic data, this work and Gaia measurements. * indicates a private communication.

| # | WDS | # obs | First Obs | Last Obs | PA (deg) | PA SEM (deg) | ρ (arcsec) | ρ SEM (arcsec) |
|---|---|---|---|---|---|---|---|---|
| 1 | 06018-4110 | 13 | 1837 | 1999 | 240.49 | 0.83 | 23.68 | 0.25 |
| 2 | 06387-4504 | 6 | 1913 | 1991 | 26.23 | 1.00 | 1.90 | 0.28 |
| 3 | 06387-4504 | 17 | 1835 | 1999 | 331.40 | 1.77 | 17.96 | 0.30 |
| 4 | 06393-3150 | 10 | 1911 | 1999 | 27.89 | 0.89 | 23.77 | 0.13 |
| 5 | 06443-3219 | 13 | 1911 | 1999 | 264.39 | 0.81 | 4.77 | 0.19 |
| 6 | 07289-3151 | 29 | 1835 | 1999 | 53.81 | 1.26 | 8.68 | 0.44 |
| 7 | 07341-5050 | 11 | 1836 | 1999 | 219.78 | 0.80 | 47.69 | 0.23 |
| 8 | 08057-3334 | 16 | 1837 | 1999 | 88.32 | 1.15 | 22.06 | 0.37 |
| 9 | 08057-3334 | 5 | 1903 | 2015* | 60.08 | 0.69 | 27.84 | 0.20 |
| 10 | 08057-3334 | 6 | 1897 | 1999 | 9.30 | 0.74 | 14.03 | 0.24 |

## 4.0 Gaia Measurements

Gaia position and proper motion measurements were sourced from the SIMBAD astronomical database and *Aladin Sky Atlas*. Gaia *PA* and *ρ* measures presented in Table 3 have been precessed to epoch of equinox J2000.0 for comparison with the observation presented in this work and historic extrapolations (Figure 1), and to estimate accuracy using the micro-arcsecond precision of the Gaia position measurements as a standard for comparison.

Table 3: *PA* and *ρ* based on Gaia data at epoch of observation.

| # | Comp | Gaia ID | Gaia data release | PA (deg) | PA δ (deg) | ρ (arcsec) | ρ δ (arcsec) |
|---|---|---|---|---|---|---|---|
| 1 | A | 2882343210792764544 | Gaia EDR3 | 240.584 | 0.007 | 23.665 | 0.015 |
| | B | 2882343210794006656 | | | | | |
| | C | 2882343107714792448 | | | | | |
| 2 | A | 5556069272925158656 | Gaia DR2 | 25.137 | <0.001 | 1.897 | 0.224 |
| | B | 5556069272923110000 | | | | | |
| 3 | A | 5556069272925158656 | Gaia DR2 | 331.375 | 0.010 | 18.007 | 0.020 |
| | C | 5556069479083586688 | | | | | |
| 4 | A | 5583873241933194496 | Gaia EDR3 | 27.936 | 0.070 | 23.751 | 0.031 |
| | B | 5583873310652668160 | | | | | |
| | C | 5583873310651519104 | | | | | |
| 5 | A | 5583926636967008384 | Gaia DR2 | 264.324 | 0.024 | 4.801 | 0.123 |
| | B | 5583926735748050944 | | | | | |
| 6 | A | 5593011729755869696 | Gaia DR2 | 53.700 | 0.025 | 8.923 | 0.041 |
| | B | 5593011832835083008 | | | | | |
| 7 | A | 5493532792450982144 | Gaia DR2 | 219.817 | 0.001 | 48.376 | 0.014 |
| | B | 5493532693670905088 | | | | | |
| 8 | A | 5546088112536703488 | Gaia DR2 | 88.187 | 0.008 | 22.071 | 0.010 |
| | B | 5546088112527125632 | | | | | |
| 9 | A | 5546088112536703488 | Gaia DR2 | 59.904 | 0.006 | 27.903 | 0.010 |
| | B | 5546088112536693376 | | | | | |
| 10 | B | 5546088112527125632 | Gaia DR2 | 8.897 | 0.054 | 13.456 | 0.175 |
| | C | 5546088112536693376 | | | | | |

## 5.0 Measurement Bias.

Figure 1 presents the differences in *PA* and *ρ* measurements from this work, Gaia DR2 and EDR3, and historical data; all at J2000.0. The largest discrepancies were caused by 06387-4504 HJ 3882AB due to the small separations involved, and 06018-4110 HJ 3827AB,C with a large delta magnitude. These pairs where removed from Figure 1, however are present in Table 1.

Table 4 shows the mean difference between the observations in this work, Gaia, and historic data (column 1). Columns 2 and 3 are the mean differences of *PA* and *ρ*. Columns 4 and 5 are the same but with 06387-4504 HJ 3882AB measurements removed from the calculation.

Table 4: Accuracy of measurements

| | PA (deg) | ρ (arcsec) | PA (deg) | ρ (arcsec) |
|---|---|---|---|---|
| Gaia – This Work | -0.430 | 0.011 | -0.137 | 0.003 |
| Hist – This Work | -0.248 | -0.036 | -0.063 | -0.045 |
| Gaia – Hist | -0.169 | 0.039 | -0.060 | 0.040 |

Figure 1: Plot of the difference in measures at epoch 2021.066. Gaia - This Work (X), Hist - This Work (□), Gaia - Hist (△).

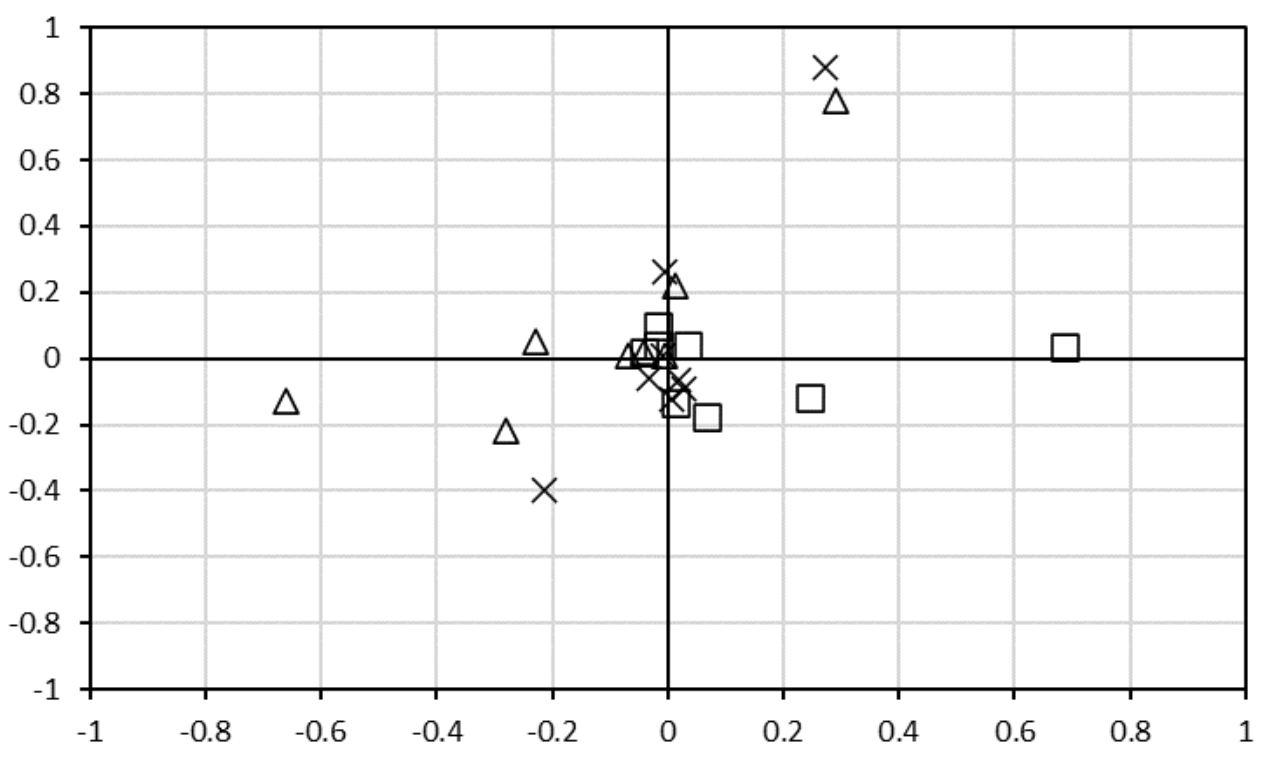

**Summary.**

Presented are *PA* and *ρ* measures of 10 scarcely observed pairs. Additionally, a comparison with Gaia DR2 & EDR3 data, and linear extrapolations of historic measurements was conducted as a check on the bias in this work.

**Acknowledgement.**

- The University of New England (UNE) for use of the Kirby observatory.
- The Aladin sky atlas developed at CDS, Strasbourg Observatory, France, https://aladin.u-strasbg.fr/
- The Washington Double Star Catalog maintained by the UNSO. (WDS), https://ad.usno.navy.mil/wds
- This work has made use of data from the European Space Agency (ESA) mission Gaia (https://www.cosmos.esa.int/gaia), processed by the Gaia Data Processing and Analysis Consortium (DPAC, https://www.cosmos.esa.int/web/gaia/dpac/consortium). Funding for the DPAC has been provided by national institutions, in particular the institutions participating in the Gaia Multilateral Agreement.
- Thanks to Florent Losse for use of the Reduc software.

**References.**